\documentclass[prd,twocolumn,showpacs,tightenlines,amssymb,eqsecnum,
               superscriptaddress,floatfix]{revtex4}
\usepackage{epsfig,latexsym,hyperref}

\begin{document}
{\flushleft{\small{\em Phys. Rev. D {\bf 66}, 064019 (2002).}}}
\title{Gravitational Waves from a Fissioning White Hole}

\author{Roberto G\'omez}
\email{gomez@psc.edu}
\affiliation{Pittsburgh Supercomputing Center,
             4400 Fifth Avenue, Pittsburgh, Pennsylvania 15213}
\affiliation{Department of Physics and Astronomy,
             University of Pittsburgh, Pittsburgh, Pennsylvania 15260}

\author{Sascha Husa}
\affiliation{Albert Einstein Institute, Max Planck Gesellschaft,
             Haus 1, Am M\"uhlenberg, Golm, Germany}
\affiliation{Department of Physics and Astronomy,
             University of Pittsburgh, Pittsburgh, Pennsylvania 15260}

\author{Luis Lehner}
\affiliation{Department of Physics and Astronomy,
         University of British Columbia, Vancouver, British Columbia,
         Canada V6T 1Z1}

\author{Jeffrey Winicour}
\affiliation{Albert Einstein Institute, Max Planck Gesellschaft,
             Haus 1, Am M\"uhlenberg, Golm, Germany}
\affiliation{Department of Physics and Astronomy,
             University of Pittsburgh, Pittsburgh, Pennsylvania 15260}

\date{9 May 2002}

\begin{abstract}

We present a fully nonlinear calculation of the waveform of the
gravitational radiation emitted in the fission of a vacuum white
hole. At early times, the waveforms agree with close approximation
perturbative calculations but they reveal dramatic time and
angular dependence in the nonlinear regime. The results pave the
way for a subsequent computation of the radiation emitted after a
binary black hole merger.

\end{abstract}

%\keywords{waveforms, white hole fission, black hole mergers}
\preprint{AEI-2002-0XX}

\maketitle

\section{Introduction}

The omputation of gravitational radiation from the inspiral and
merger of binary black holes poses a difficult boundary value
problem. In the geometrically simplest and physically most natural
treatment, the black holes are modeled by the gravitational
collapse of a pair of stars (or other astrophysical bodies).
However, this is a challenging hydrodynamic problem which requires
simulating a pair of orbiting bodies for a sufficient time to
verify a negligible amount of incoming radiation in the initial
conditions, then following their subsequent collapse to black
holes and finally computing the outgoing radiation in the exterior
spacetime. Alternatively, in the purely vacuum approach, the
individual black holes form from imploding gravitational waves.
This avoids hydrodynamical difficulties at the expense of a
globally complicated initial value problem. The imploding waves
may emanate either (i) from a past singularity or (ii) from past
null infinity ${\cal I}^-$. In case (i), the appropriate boundary
condition at ${\cal I}^-$ is that there be no ingoing radiation
but, assuming the time reversed version of cosmic censorship, the
past singularity implies a white hole horizon ${\cal H}^-$ on
which boundary data must be specified in some arbitrary manner in
order to determine the exterior spacetime. In case (ii), ingoing
radiation from ${\cal I}^-$ is present at early times when the
black holes are formed but ingoing radiation must be absent at
late times in order for the outgoing radiation to be unambiguously
attributed to the merging black holes. In this work, we present a
solution to the first stage of a new two-stage global treatment of
the vacuum binary black hole problem~\cite{kyoto,close1}. The
approach, based upon characteristic evolution, has been carried
out in the regime of Schwarzschild perturbations where advanced
and retarded solutions of the linearized problem can be rigorously
identified~\cite{close2}. Computational experiments are necessary
to study the applicability of the approach to the nonlinear
regime.

From a time-reversed viewpoint, this first stage is equivalent to
the determination of the outgoing radiation emitted from the
fission of a white hole in the absence of ingoing radiation. This
provides the physically correct ``retarded'' waveform for a white
hole fission, were such events to occur in the universe. Although
there is no standard astrophysical mechanism for producing white
holes from a nonsingular matter distribution, white holes of
primordial or quantum gravitational origin cannot be ruled out.

This fission problem has a simpler formulation as a characteristic
initial value problem than the black hole merger problem. The
boundary of the (conformally compactified) exterior spacetime
contains two null hypersurfaces where boundary conditions must be
satisfied: past null infinity ${\cal I}^-$, where the incoming
radiation must vanish, and the white hole event horizon ${\cal
H}^-$, which must describe a white hole, which is initially in
equilibrium with no ingoing radiation and then distorts and
ultimately fissions into two white holes with the emission of
outgoing gravitational waves.  If we approximate  ${\cal I}^-$ by
an outgoing null hypersurface $J^-$, which intersects ${\cal H}^-$
at an early time (approximating past time infinity $i^-$) close to
the initial equilibrium of the white hole, then data on these two
null hypersurfaces, ${\cal H}^-$ and $J^-$, constitute a standard
double-null initial value problem, whose evolution determines a
portion of the exterior spacetime extending to ${\cal I}^+$, where
the outgoing radiation is computed. In contrast, the corresponding
problem for the ``retarded'' waveform from a black hole merger
involves two disjoint null hypersurfaces where boundary conditions
must be satisfied: past null infinity ${\cal I}^-$, where the
incoming radiation must vanish, and the future event horizon
${\cal H}^+$, which describes the merger of the two black holes
and their subsequent approach to equilibrium.

In previous work~\cite{close1}, we treated the fission problem in
the close approximation~\cite{pp} as a perturbation of a
Schwarzschild background. In this paper we present a fully
nonlinear treatment that reveals new and interesting strong field
behavior. We carry out the evolution of this vacuum double-null
problem by means of a characteristic evolution
code~\cite{cce,news,wobb}, using a recent version of the code
which improves accuracy in the highly nonlinear
region~\cite{reduced}. Caustics in the ingoing null hypersurfaces
used to foliate the exterior spacetime restrict the evolution to
the pre-fission stage.

We use a conformal horizon model~\cite{ndata,asym,nulltube} to
supply the necessary null data for a horizon corresponding to a
white hole fission. The conformal horizon model provides a
stand-alone description of the intrinsic null geometry of the
horizon. The algorithm for generating horizon data is constructed
to handle a general event horizon representing the fission of a
spinning white hole into two outspiraling white holes of non-equal
mass~\cite{nulltube}. The specific application in this paper is to
the axisymmetric head-on fission into equal mass white holes. (The
necessary data and evolution codes are, however, {\it not} limited
to the axial symmetry of a head-on collision.) The resulting
horizon geometry is an upside-down version of the standard
trousers-shaped event horizon for a binary black hole merger in
the time-reversed scenario.

We study a range of models extending from the perturbative close
limit, in which the fission occurs in the infinite future, to the
highly nonlinear regime. Nontrivial global changes, accompanied by
dramatic time dependence of the horizon geometry, arise in passing
from the perturbative to the highly nonlinear
regime~\cite{nulltube}. The existence of a marginally trapped
surface divides the horizon into interior and exterior regions,
analogous to the division of the Schwarzschild horizon by the
$r=2M$ bifurcation sphere. In passing from the perturbative to the
strongly nonlinear regime there is a transition in which the
fission occurs in the portion of the horizon visible from ${\cal
I}^+$. Thus these results reveal two classes of binary white hole
spacetimes, depending upon whether the crotch in the trousers
(where the fission occurs) is bare, in the sense that it is
visible from ${\cal I}^+$, or hidden inside a marginally trapped
surface. In this paper we evolve this data using the
characteristic code to study the properties of the gravitational
radiation produced by this dramatic behavior of a white hole
fission.

In Secs.~\ref{sec:data} and \ref{sec:confmod} we review the
formalism and data necessary for a characteristic evolution of the
spacetime exterior to a dynamic white hole by means of the
characteristic code. In Sec.~\ref{sec:wave} we present the
detailed waveforms for a one-parameter family of spacetimes
varying from the close approximation to the highly nonlinear
regime in which the fission is visible from ${\cal I}^+$.

We retain the conventions of our previous
papers~\cite{cce,news,wobb,ndata,asym}, with only minor changes
where noted in the text. For brevity, we use the notation
$f_{,x}=\partial_x f$ to denote partial derivatives and $\dot f
=\partial_u f$ to denote retarded time derivatives. We represent
tensor fields on the sphere as spin-weighted
variables~\cite{penrin} in terms of a dyad $q_A$ for the unit
sphere metric $q_{AB}=q_{(A}q_{B)}$ in some standard choice of
spherical coordinates, e.g. $x^A =(\theta,\phi)$. (The numerical
code uses two overlapping stereographic coordinate patches.) We
compute angular derivatives of tensor fields in terms of $\eth$
and $\bar\eth$ operators~\cite{eth}, e.g. $f_{,A}=\Re ( \bar q_A
\eth f)$, to compute the gradient of a spin-weight zero (scalar)
field $f$ in terms of the spin-weight $1$ field $\eth f$ and the
spin-weight $-1$ field $\bar \eth f$.

\section {The double-null problem for a fissioning white hole}
\label{sec:data}

We treat the fission of a white hole by a double null initial value
problem based upon the white hole horizon ${\cal H}^-$ and on an outgoing
null hypersurface ${\cal J}^-$, which emanates from an early time slice
${\cal S}^-$ of ${\cal H}^-$ approximating the initial equilibrium of the
white hole. The horizon pinches off in the future where its generators
either caustic or cross, producing the (upside-down) trousers picture
of a fissioning white hole.

The double-null problem, first formulated by Sachs~\cite{sachsdn},
is most conveniently described in Sachs coordinates consisting of
(i) an affine null parameter $u$ along  the generators of ${\cal
H}^-$, which foliates ${\cal H}^-$ into cross sections ${\cal
S}_u$ and labels the corresponding outgoing null hypersurfaces
${\cal J}_u$ emanating from the foliation, (ii) angular
coordinates $x^A$ which are constant both along the generators of
${\cal H}^-$ and along the outgoing null rays of ${\cal J}_u$, and
(iii) an affine parameter $\lambda$ along the outgoing rays
normalized by $\nabla^{\alpha}u \nabla_{\alpha}\lambda =-1$, with
$\lambda =0$ on ${\cal H}^-$.  In these $x^{\alpha}=(u,\lambda
,x^A)$ Sachs coordinates, the metric takes the form
\begin{eqnarray}
   ds^2  &=& -(W -g_{AB}W^A W^B)du^2 -2dud\lambda \nonumber \\
   &-& 2g_{AB}W^Bdudx^A +  g_{AB}dx^Adx^B
\label{eq:amet}
\end{eqnarray}
We set $g_{AB}=r^2h_{AB}$, where $\det(h_{AB})=\det(q_{AB})=q(x^A)$,
with $q_{AB}$ the unit sphere metric. We represent the
conformal metric $h_{AB}$ by the complex spin-weight 2 field
$J=\frac{1}{2}h_{AB}q^Aq^B$. The remaining dyad component, given by the
real function $K=\frac{1}{2}h_{AB}q^A\bar q^B$, is fixed by the
determinant condition $K^2=1+J\bar J$.

The requirement that the horizon be null implies that $W=0$ on ${\cal
H}^-$. In addition, we fix the gauge freedom corresponding to the
shift on ${\cal H}^-$ so that $n^a \partial_a =\partial_u$ is tangent
to the generators, implying that $W^A=0$ on ${\cal H}^-$. The choice of
lapse that $u$ is an affine parameter implies $\partial_\lambda W=0$ on
${\cal H}^-$.  We further fix the affine freedom by specifying $u=u_-$
on the early slice ${\cal S}^-$ approximating the asymptotic equilibrium
of the white hole in the past. The outgoing null hypersurface ${\cal J}^-$
emanating from ${\cal S}^-$ approximates past null infinity ${\cal I}^-$.

The affine tangent to the generators of ${\cal H}^-$,
$n^a\partial_a=\partial_u$ satisfies the geodesic equation
$n^b\nabla_b n^a=0$ and the hypersurface orthogonality condition
$n^{[a}\nabla^b n^{c]}=0$.  Following the approach of
Refs.~\cite{ndata,asym}, we project four-dimensional tensor fields
into the tangent space of ${\cal H}^-$ using the operator

\begin{equation}
       P_a^b = \delta_a^b + n_a l^b,
\end{equation}
where $l_a = -\nabla_a u$.

The evolution proceeds along the outgoing null hypersurfaces ${\cal
J}_u$ emanating from the foliation of ${\cal H}^-$ and extending
to (compactified) ${\cal I}^+$.  In this problem, the complete (and
unconstrained) characteristic data on ${\cal H}^-$ are its (degenerate)
intrinsic conformal metric $h_{AB}(u,x^A)$, or equivalently $J(u,x^A)$,
expressed in terms of the affine parameter $u$.  In addition, the
characteristic data on ${\cal J}^-$ are its intrinsic conformal metric
$h_{AB}(\lambda,x^A)$, or  $J(\lambda,x^A)$ expressed in terms of its
affine parameter $\lambda$.

The remaining data necessary to evolve the exterior spacetime consist of
the intrinsic metric and extrinsic curvature of ${\cal S}^-$ (subject
to consistency with the characteristic data)~\cite{sachsdn,haywdn}.
This additional data consist of the surface area $r$, the inward expansion
$\dot r$, the outward expansion $r_{,\lambda}$ and the twist~\cite{haywsn}

\begin{equation}
       \omega_a =P^b_a n^c \nabla_c l_b =(0,0,\omega_A) .
\end{equation}
(Our choice of shift implies that $\omega_u=\omega_{\lambda}=0$.)
The twist is an invariantly defined extrinsic curvature property
of the $u={\rm const}$ cross sections of ${\cal H}^-$, independent
of the boost freedom in the extensions of $n_a$ and $l_a$ subject
to the normalization $n^a l_a=-1$.

We use the conformal horizon model to supply data $J$ on ${\cal H}^-$
corresponding to a fissioning white hole for a sequence of models ranging
from the close approximation to the highly nonlinear regime (see Sec.
\ref{sec:confmod}). On ${\cal J}^-$, we set $J=0$ to model the absence
of ingoing radiation. In carrying out the evolution computationally, the
first step is to propagate the data on ${\cal S}^-$ along the generators
of ${\cal H}$~\cite{sachseq} so that it can be supplied as boundary data
for the exterior characteristic evolution code.

\subsection{Propagation equations on the horizon}
\label{sec:horprop}

Einstein's equations for the double-null problem decompose into (i)
hypersurface equations intrinsic to the null hypersurfaces ${\cal J}_u$,
which determine auxiliary metric quantities in terms of the conformal
metric $h_{AB}$; (ii) evolution equations which determine the rate of
change $\partial_u h_{AB}$ of the conformal metric of ${\cal J}_u$;
and (iii) propagation equations which are constraints that need only be
satisfied on ${\cal H}^-$~\cite{sachsdn}. The Bianchi identities ensure
that the propagation equations will be satisfied in the exterior spacetime
as a result of the hypersurface and evolution equations. Integration
of the propagation equations determines the additional horizon data
necessary for the characteristic evolution.

One of the propagation equations is the ingoing Raychaudhuri
equation $R_{uu}=0$, which propagates the surface area variable
$r$ along the generators of ${\cal H}^-$ in terms of initial
conditions on ${\cal S}^-$. The value of $\dot r$ on ${\cal H}^-$
determines the convergence of the ingoing null rays. Once the
intrinsic geometry $J$ and the area coordinate $r$ are known, the
vacuum equation $R_{Au}=0$ is used to propagate the twist
$\omega=q^A \omega_A$ along the generators of ${\cal H}^-$. This
determines $\omega$ in terms of its initial value at $S^{-}$. The
$R_{AB}=0$ vacuum equations propagate the outward expansion and
shear of the foliation ${\cal S}_u$ along ${\cal H}^-$. The trace
part propagates the outgoing expansion determined by
$r_{,\lambda}$. The trace-free part is an evolution equation for
$h_{AB,\lambda}$, which describes the shear of the outgoing rays.

In summary, the data for the double null problem include the
conformal metric $h_{AB}$ on ${\cal H}^-$ and ${\cal J}^-$ and the
quantities $r$, $\dot r$, $\omega$ and $r_{,\lambda}$ on ${\cal
S}^-$. Equations~(A1), (2.29), (2.30), and (2.34) of
Ref.~\cite{nulltube} propagate the data on ${\cal S}^-$ to all of
${\cal H}^-$. The data required on ${\cal S}^-$ can be inferred
from the asymptotic properties of the white hole equilibrium at
$i^-$. The propagation equations, given in spin-weighted form
in~\cite{nulltube}, are implemented numerically with a second
order Runge-Kutta scheme~\cite{recipes}. Complete details can be
found in Ref.~\cite{nulltube}.

\subsection{Bondi-Sachs coordinates and characteristic code variables}
\label{sec:Bondisachs}

The null code is based upon the Bondi-Sachs version of the
characteristic initial value problem~\cite{bondi,sachs}. It is
designed to evolve forward in time along a foliation of spacetime
by outgoing null hypersurfaces. The Bondi-Sachs coordinates differ
from the Sachs coordinates (\ref{eq:amet}) by the use of a surface
area coordinate $r$ along the outgoing cones rather than the
affine parameter $\lambda$. Because a generic horizon is not a
hypersurface of constant $r$, it is advantageous to first discuss
the necessary data in terms of Sachs coordinates and then
transform to the $r$ coordinate. In Bondi-Sachs variables, the
metric takes the form

%\begin{widetext}
%\begin{equation}
%     ds^2 = -\left(e^{2\beta}{V \over r} -r^2h_{AB}U^AU^B\right) du^2
%           - 2e^{2\beta}dudr -2r^2 h_{AB}U^Bdudx^A + r^2h_{AB}dx^Adx^B.
%     \label{eq:umet}
%\end{equation}
%\end{widetext}
\begin{eqnarray}
     ds^2 = &-\left(e^{2\beta}{V \over r} -r^2h_{AB}U^AU^B\right) du^2
           - 2e^{2\beta}dudr \nonumber \\&-2r^2 h_{AB}U^Bdudx^A + r^2h_{AB}dx^Adx^B.
     \label{eq:umet}
\end{eqnarray}
The field $U^A$ is represented in spin-weighted form $U=q_A U^A$.

The evolution of the exterior spacetime requires a transformation
of the data on ${\cal H}^-$ from Sachs coordinates
$(u,\lambda,x^a)$ to Bondi-Sachs coordinates $(u,r,x^a)$, as
described in detail in Ref.~\cite{nulltube}. Since $r$ is not in
general constant on ${\cal H}^-$, the horizon does not lie
precisely on radial grid points $r_i$. Consequently,  an accurate
prescription of boundary values on the $r_i$ grid points nearest
the horizon requires a Taylor expansion of the horizon data.
Restricted to ${\cal H}^-$, the metric variables $r$ and $J$ have
the same values in both Sachs and Bondi-Sachs coordinates and our
choices of lapse and shift imply that the Bondi-Sachs variables
$\beta$, $U$ and $V$ are related to Sachs variables by
Eqs.~(2.15), (2.31), and (2.32) of Ref.~\cite{nulltube}.

A similar construction~\cite{nulltube} provides their first $r$
derivatives in terms of known Sachs variables. We obtain $J_{,r}$
from Eq.~(2.33) of Ref.~\cite{nulltube}, $\beta_{,r}$ from the
$R_{\lambda\lambda}=0$ Raychaudhuri equation for the outgoing null
geodesics, Eq.\,(2.37) of Ref.~\cite{nulltube}, and the value of
$\partial_r U$ on the horizon from the auxiliary field $Q=q^A
Q_A$, where

\begin{equation}
     Q_A = r^2 e^{-2\,\beta} h_{AB} U^B_{,r}.
\end{equation}
Here $Q$ is obtained from the twist and other Sachs variables as
per Eq.~(2.39) of~\cite{nulltube}. Finally, we compute $\partial_r
V$ by obtaining  $ V_{,\lambda}$ from the $\lambda$ derivative of
Eq.~(2.32) of Ref.~\cite{nulltube}.

The values of each metric function $(J,\beta,U,V)$ and its first radial
derivative can be used to consistently initialize field values at
the $r_i$-grid points near the horizon. Boundary values for the code are
then provided, to second order accuracy, on grid points bracketing the
horizon. For each outgoing null ray the value of the area coordinate
$r_{{\cal H}^-}$ on the horizon is known. This value is bracketed by
the nearest grid points $r_{i-1} \le r_{{\cal H}^-} < r_i$, where the
metric values are computed by the second order accurate Taylor expansion
\begin{equation}
   J_{r_i}= J|_{{\cal H}^-} + \left( r_i - r_{{\cal H}^-} \right)
            J_{,r}|_{{\cal H}^-}.
\end{equation}
After the metric quantities have been computed at the grid points
neighboring the horizon, the code can evolve the spacetime exterior to
the horizon.

\subsection{Evolution equations}
\label{sec:evolve}

The system of equations that determines the exterior spacetime
forms a hierarchy~\cite{newt,nullinf}. Alternative formulations
are available; see for instance Ref.~\cite{news}. Here we use a
recent formulation~\cite{reduced}, which was specifically
developed to handle the extremely nonlinear post-merger regime of
binary black hole collisions. It reduces all angular derivatives
to first order by introducing the auxiliary variables $\nu=\eth
J$, $k=\eth K$ and $B=\eth \beta$. This results in the hierarchy
of hypersurface equations and one complex evolution equation for
the conformal metric function $J$ given by Eqs.~(16)--(25) of
Ref.~\cite{reduced}. We note that the terms on the right hand side
of the evolution equation, Eq.~(16) of~\cite{reduced}, have been
grouped into hypersurface terms $J_H$ which vanish for linear
perturbations of a spherically symmetric spacetime and a term
$P_u$ which isolates the only nonlinear term containing a
(retarded) time derivative of $J$. The introduction of the
auxiliary variables $B$, $\nu$, and $k$ in ~\cite{reduced}
eliminates all second angular derivatives from the hypersurface
and evolution equations. This leads to substantially improved
numerical behavior~\cite{reduced} over the standard characteristic
formulation~\cite{cce,news,matter,stable} in the extremely
nonlinear post-merger regime of binary black hole collisions.
Details are given in Ref.~\cite{reduced}.

We retain the radial and time integration schemes of the standard
characteristic formulation. The radial integration algorithm is
explained in detail in Refs.~\cite{axisymmetric,cce,news,disip}.
In a departure from Ref.~\cite{news}, we use a three-step
iterative Crank-Nicholson scheme~\cite{sfrg} to ensure stability
of the time evolution~\cite{teuk}.

The evolution code first integrates the hypersurface equations
[Eqs.~(17)--(23) of~\cite{reduced}] on the initial null
hypersurface ${\cal J}^-$ in the region exterior to the horizon.
The evolution equation, Eq.~(16) of~\cite{reduced} is then used to
advance to the next hypersurface in the region exterior to the
horizon, etc., advancing the computation of the exterior spacetime
in the region from the horizon to ${\cal I}^+$. The evolution
continues as long as the coordinate system remains well behaved.

\subsection{The Bondi news}
\label{sec:news}

The Bondi news function $N(u,x^B)$~\cite{bondi,Penrose,high} is an
invariantly defined field on ${\cal I}^+$ which gives the amplitude
of the flux of gravitational wave energy. Its computation requires
evaluation of the conformal factor $\Omega_{conf}$ necessary to
compactify the spacetime in a conformal Bondi frame~\cite{tam}, which
is an asymptotic inertial frame in which the slices of ${\cal I}^+$
have unit sphere geometry. Following the approach of Ref.~\cite{news},
we set  $\Omega_{conf} =\omega_{conf}/r$, where $\omega_{conf}(u,x^A)$
is a smooth non-vanishing field at ${\cal I}^+$. At the early time $u_-$
at which we initiate the evolution, the conformal metric $h_{AB}$ of the
outgoing null hypersurface approaches the unit sphere metric $q_{AB}$
at ${\cal I}^+$ so that $\omega_{conf-}=\omega_{conf}|_{u_-}=1$.
However, as the evolution proceeds, the $(u,x^A)$ coordinates, which
are naturally adapted to ${\cal H}^-$, become non-inertial at ${\cal
I}^+$. This results in the time dependence

\begin{equation}
     (\partial_u +L^A \partial_A )\log \omega_{conf}=
           - \frac{1}{2} D_A L^A
\label{eq:conf}
\end{equation}
where $L^A$ is the asymptotic value of $U^A$ at ${\cal
I}^+$~\cite{news}.

In addition, a transformation to conformal Bondi coordinates $(u_B,y^A)$
on ${\cal I}^+$ is necessary to determine the Bondi news

\begin{equation}
      {\cal N}(u_B,y^A) =N\big( u(u_B,y^A), x^A(u_B,y^A) \big ) ,
\end{equation}
as measured in an inertial frame. The relevant expressions are
given in Sec.  IV and Appendix~B of Ref.~\cite{news}. Inspection
of Eqs.~(B1)--(B6) of Ref.~\cite{news} reveals that second angular
derivatives of the conformal factor $\omega_{conf}$ enter the
calculation. For improved accuracy, we remove these second
derivatives by introducing~\cite{reduced} the auxiliary variable
${\cal W}=\eth\omega_{conf}$. Since $\omega_{conf}$ is defined
solely on ${\cal I}^{+}$, we use the consistency relation ${\cal
W}_{,u}=\eth\omega_{conf,u}$ to propagate ${\cal W}$ along the
generators of ${\cal I}^{+}$, initializing it by ${\cal W}_-
=\eth\omega_{conf-}$. We use a combination of second-order
Runge-Kutta and mid-point rule schemes~\cite{reduced} for the time
integration of Eq.~(\ref{eq:conf}).

\section{Null data for the axisymmetric head-on fission}
\label{sec:confmod}

The conformal horizon model~\cite{ndata,asym} supplies the
conformal metric $h_{AB}$ constituting the null data for a binary
black or white hole. For the case of a head-on fission of a white
hole, the  model is based upon the flat space null hypersurface
${\cal H}$ emanating from a prolate spheroid ${\cal S}_0$ of
eccentricity $\epsilon$ and semimajor axis $a$, which is embedded
at a constant inertial time $\hat t=0$ in Minkowski space. Traced
back into the past, ${\cal H}$ expands to an asymptotically
spherical shape. Traced into the future, ${\cal H}$ pinches off at
points where its null rays cross or at caustic points where
neighboring null rays focus.

With the dyad choice $q^A=(1,i/\sin\theta)$, the intrinsic conformal
metric $h_{AB}$ of ${\cal H}$ as a flat space null hypersurface is
determined by the spin-weight-2 field

\begin{eqnarray}
  J&=&{1\over 2}q^A q^B h_{AB}(\hat t,x^C) \nonumber \\
   &=& \frac{1}{2}\bigg ( \frac { \hat t -r_\theta}
                       {\hat t -r_\phi} \bigg )
                    -\frac{1}{2}\bigg ( \frac {\hat t -r_\phi}
                       { \hat t -r_\theta} \bigg ),
\label{eq:headonj}
\end{eqnarray}
where $r_\theta$ and $r_\phi$ are the principle radii of curvature of
the spheroid. The metric of the spheroid induced by its embedding in a
flat space is  $\hat g_{AB}=\hat r^2 h_{AB}$, where $\hat r^2\sin\theta$
is the local surface area.

The white hole horizon shares the same manifold ${\cal H}$ and the
same (degenerate) conformal metric as its Minkowski space
counterpart and Eq.~(\ref{eq:headonj})  provides the conformal
null data for the white hole horizon in the $\hat t$ foliation.
But the surface area and affine parametrization of the white hole
horizon and its flat space counterpart differ. As a white hole
horizon, ${\cal H}$ extends infinitely far to the past of ${\cal
S}_0$ to an asymptotic equilibrium with finite surface area. The
intrinsic metric of the white hole horizon is given by
$g_{AB}=\Omega^2 \hat g_{AB}$, and has local surface area
$r^2\sin\theta$ where $r=\Omega \hat r$. The conformal factor
$\Omega$ is designed to stop the expansion of the white hole in
the past so that the surface area radius asymptotically hovers at
a fixed value $r=R_\infty$. The conformal factor is given
by~\cite{ndata,asym}

\begin{equation}
    \Omega=-R_{\infty}\Big( {\hat u
          +\frac{\sigma^2}{12(p-\hat u )}} \Big)^{-1},
\label{eq:ansatz}
\end{equation}
where $p$ is a model parameter, $\sigma =r_{\theta} - r_{\phi}$ is
the difference between the principal curvature radii, and $\hat u$ is
a flat space affine parameter along the generators of ${\cal H}$ with
the same scale as $\hat t$ but with its origin shifted along each ray by
$\hat u =\hat t -\frac {1}{2}(r_{\theta} + r_{\phi})$. For an initially
Schwarzschild white hole of mass $M$, $R_{\infty}=2M$. Smoothness of
the white hole requires that the parameter $p \ge \sigma_M/\sqrt{13}$,
where $\sigma_M$ is the maximum value of $\sigma$ attained on ${\cal
S}_0$. (For the prolate spheroid considered here, $\sigma_M =a\epsilon$.)

Both as a flat space null hypersurface and as a white hole
horizon, ${\cal H}$ must obey the Raychaudhuri equation, which
governs the second derivative of the surface area with respect to
the affine parameter. Because the focusing power determined by
their conformal geometries are the same, the Raychaudhuri equation
implies

\begin{equation}
  \frac {1}{r} \partial_t^2 r =\frac {1}{\hat r} \partial_{\hat t}^2 \hat r.
\end{equation}
Consequently, the different behavior of the surface area
coordinates $r$ and $\hat r$ due to the conformal factor $\Omega$
implies that the affine parametrization $t$ on the white hole
horizon is related to its flat space counterpart $\hat u$
according to~\cite{ndata}

\begin{eqnarray}
  \frac {dt}{d \hat t} &=& \frac{9}{(12 \hat u (\hat u-p) -
      \sigma^2)^2}  \nonumber \\
  & \times &
  \frac{( 5 p + \mu-2 \hat u)^{2 \, (2p/\mu +1) \,}}
       {( 5 p - \mu-2 \hat u)^{2 \, (2p/\mu -1) \,}} \, ,
\label{eq:lampr}
\end{eqnarray}
where
\begin{equation}
     \mu = \sqrt{13p^2 -\sigma^2}
\label{eq:def_mu}
\end{equation}
and where the affine scale is fixed by the condition $dt/ d\hat t
\rightarrow 1$ as $\hat t \rightarrow -\infty $. The scale of the
particular affine parameter $u$ used for the time step in the
evolution code is related to $t$ by $u=t(1+\epsilon \cos^2
\theta)^{3/2}$ (chosen so that the null data $J$ has early time
behavior of the pure spin-weight 2 quadrupole form $J\sim
-\epsilon \sin^2\theta /u$).

Equation (\ref{eq:lampr}) determines the rate of deviation of a slicing
adapted to an affine parameter $u$ of the white hole horizon from the
$\hat t$ slicing given by the Minkowski embedding. The build up of a large
relative angular dependence between the $u$ and $\hat t$ foliations leads
to the change in topology of the fissioning white hole and the associated
pair-of-pants shaped horizon. We begin our evolution at an early slice
${\cal S}^-$, at $u=u_-$, where the horizon is close to equilibrium as
a Schwarzschild white hole. Initially, the $u$ and $\hat t$ foliations
behave quite similarly. The time at which nonlinear behavior becomes
significant is controlled by the location of the Minkowski spheroid ${\cal
S}_0$, at $\hat t=0$, with respect to the startup slice ${\cal S}^-$.

\section{Waveforms from fissioning white holes}
\label{sec:wave}

We use the characteristic code~\cite{reduced} to evolve the
exterior spacetime of an axisymmetric, head-on white hole fission
and compute the detailed waveforms for a sequence of spacetimes in
the range $0 \le \epsilon \le 10^{-2}$, varying from the
perturbative to the highly nonlinear regime. The object is to
search for new gravitational wave physics resulting from the
highly nonlinear behavior of a binary horizon. For that purpose,
the simulations reported here were carried out on a grid of
$77\times 77\times 137$ (stereographic patch $\times$ radial)
points. At this resolution, a higher eccentricity evolution
($\epsilon \approx 10^{-2}$) requires approximately one hour on a
single Pentium II 750 Mhz processor. Since the numerical code has
been demonstrated to be second order convergent, more accurate
results, especially concerning late time behavior near the point
of fission can be attained by increasing the numerical grid
resolution. Even though the characteristic code~\cite{reduced} is
not parallel, parameter search runs of the type described here are
entirely feasible on computing clusters with the present code.

We initialize the simulations according to the specifications in
Ref.~\cite{nulltube}, where the horizon geometry for this data was
computed and studied. We begin the evolution at $u=u_-=-100$ with the
initial mass $M_-=100$ and conformal model parameters $\hat t_- =-10$
(which fixes the location of the spheroid ${\cal S}_0$), $a=1$ (the
semi-major axis of the spheroid) and $p=(10^{-2}+10^{-5})/\sqrt{13}$
(just above the minimum value of $p$ allowed by regularity of the
conformal model for this range of eccentricities~\cite{nulltube}. On
${\cal J}^-$ ($u=u_-$), we set $J=0$ to model the absence of ingoing
radiation in an initially Schwarzschild white hole.

The limit $\epsilon\rightarrow 0$ yields the Schwarzschild solution
and small eccentricity corresponds to the close approximation where
the fission takes place far in the future behind a marginally trapped
surface so that it is hidden from external observers. For sufficiently
small $\epsilon$, the close approximation is valid in the entire exterior
Kruskal quadrant but for large $\epsilon$ it is valid only at very early
times. For large $\epsilon$, the fission occurs in the region of spacetime
visible from ${\cal I}^+$. For a critical value $\epsilon=\epsilon_c
\approx 10^{-4}$ there is a transition between these two regimes. For
$\epsilon <\epsilon_c$, the evolution terminates when the expansion
of the outgoing null hypersurfaces vanishes along some ray. (In the
Schwarzschild limit this occurs simultaneously on all rays on the black
hole horizon.) For $\epsilon >\epsilon_c$, the evolution extends to the
point of fission when $r\rightarrow 0$ on the equator of the white hole.

\begin{figure}
\epsfig{file=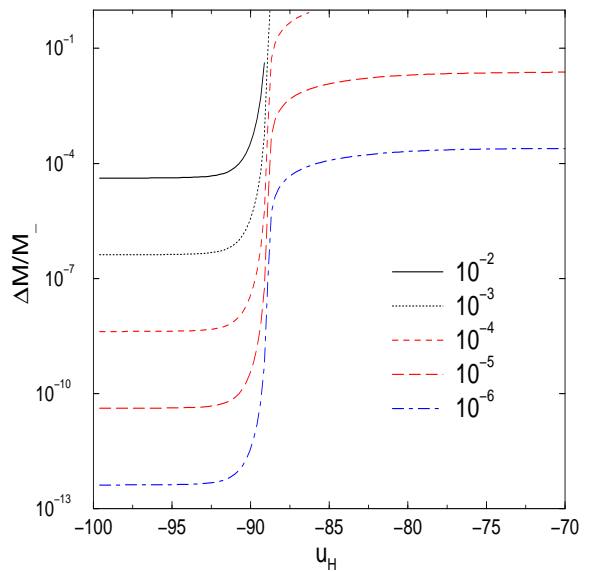,height=3.0in,width=3.0in,angle=0}
\caption{Relative mass loss $\Delta M/M_0$ as a function of the horizon
affine parameter $u$, shown here for values of the eccentricity
parameter ranging from $\epsilon=10^{-6}$ to $\epsilon=10^{-2}$.}
\label{fig:mloss}
\end{figure}

Figure \ref{fig:mloss} shows the time dependence of the fractional
mass loss $\Delta M(u)/M_- =[M(u) -M_-]/M_-$ for a sequence of
runs.
A negligible amount of mass is radiated at startup, confirming that in all
cases the initial geometry is very close to Scharzschild. The mass loss
remains small until $u\approx -90$, near the location of ${\cal S}_0$
on ${\cal H}^-$, at which time a rapid change in $d\hat{t}/du$ occurs,
and the mass loss rises sharply.

Following the sharp rise in $\Delta M/M_-$, a transient period is
observed in all the curves. The largest contribution to the radiated
energy comes from the last stage of the evolution. The total mass loss
remains well below the 1\% level until eccentricities of about $\epsilon
\approx 10^{-5}$ are reached, when the total mass loss at the end of
the simulation approaches 2\% of the initial white hole mass.

At higher eccentricities, i.e. at $\epsilon \ge 10^{-4}$, most of
the initial mass is radiated away during the later stages of the
simulation. The rate of mass loss in the final stage is
particularly dramatic in the cases $\epsilon>\epsilon_c$ where the
fission is visible from ${\cal I}^+$.

\begin{figure}
\epsfig{file=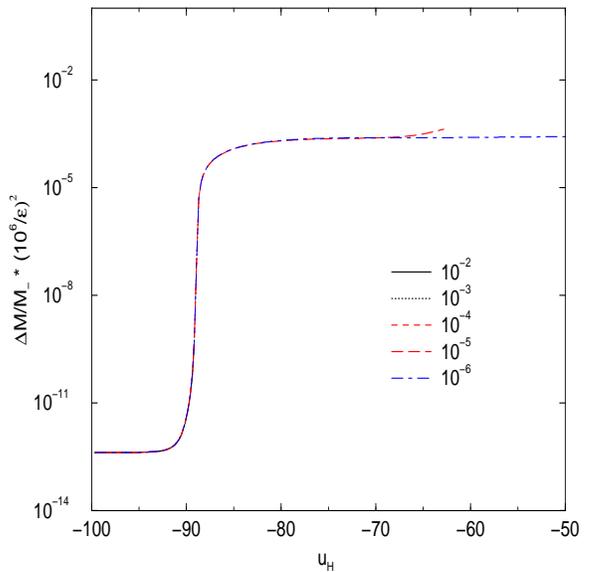,height=3.0in,width=3.0in,angle=0}
\caption{The relative mass loss $\Delta M/M_0$, scaled by $\epsilon^{2}$
and overlayed on the profile for the case $\epsilon=10^{-6}$. The striking
agreement indicates that the mass loss behavior is still $O(\epsilon^2)$.}
\label{fig:mloss_scaled}
\end{figure}

Figure~\ref{fig:mloss_scaled} shows again the mass loss profiles
of Fig.~\ref{fig:mloss}, but now scaled by the relative amplitude
$(\epsilon/\epsilon_{0})^2$, with $\epsilon_{0}=10^{-6}$. The agreement
seen clearly indicates that the behavior is $O(\epsilon^2)$ in the range
where they overlap.

\begin{figure}
\epsfig{file=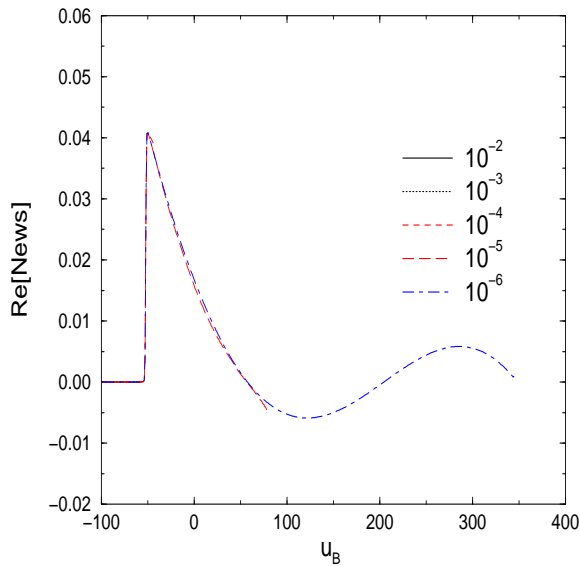,height=3.0in,width=3.0in,angle=0}
\caption{The real part of the Bondi news at a point on the equator,
scaled by $\epsilon$ and overlayed on the profile for the case
$\epsilon=10^{-6}$,
as a function of Bondi time $u_B$. A sharp pulse is clearly visible
at $u_{B}\approx -52$, which is a distinct feature of the white hole
horizon data.}
\label{fig:newseq2-6}
\end{figure}

Figure~\ref{fig:newseq2-6} shows the time dependence of the news measured
at the equator for $\epsilon=10^{-6}$  to $\epsilon=10^{-2}$.
Here we have also overlayed the plots, rescaling them by
$(\epsilon/\epsilon_{0})$, with $\epsilon_{0}=10^{-6}$. We plot the real
part of the Bondi news at a point on the equator, as a function of Bondi
time $u_B$.  Due to axisymmetry, the imaginary part of the Bondi news is
zero to within discretization error. The sharp pulse clearly visible in
the news at $u_{B}\approx -52$ results from the choice of parameters in
the white hole horizon data which control the rapidity of the fission.
Here, in order that the model yield a bare fission, the parameters have
been chosen so that the fission occurs on an extremely rapid time scale.
Subsequent to this sharp pulse, the news undergoes a damped
oscillation (which is too early in the final ringdown to be associated
with a quadrupole quasinormal mode).

The Bondi news as computed by the full code osculates the results
obtained in the perturbative limit~\cite{yosef}, in the regime in
which the perturbative and nonlinear codes are clearly solving the
same problem (up to $u_B \approx 100$). At later times ($u_B
\approx 105$), the fully nonlinear news calculation deviates from
the perturbative results as a consequence of the appearance of
higher harmonics. These higher harmonics, which are clearly
visible in the news as a function on the sphere, arise from the
nonlinearity of the equations and cannot be observed in a
perturbative evolution.

\begin{figure}
\epsfig{file=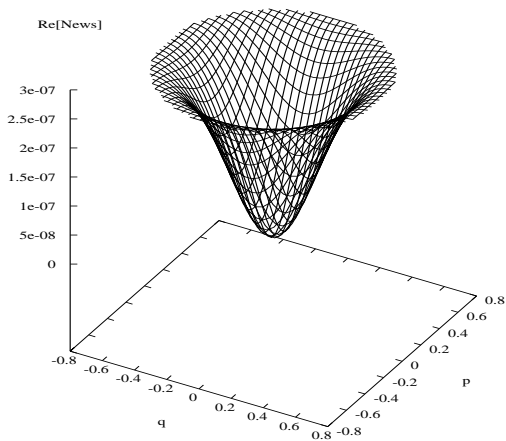,height=3.0in,width=3.0in,angle=0}
\caption{The real part of the Bondi news at early times ($u=-96.63$)
for $\epsilon=10^{-6}$. At this early time, the angular dependence is
purely quadrupolar.}
\label{fig:quad}
\end{figure}

\begin{figure}
\epsfig{file=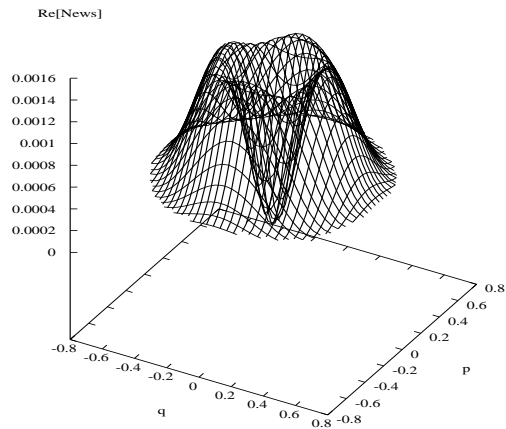,height=3.0in,width=3.0in,angle=0}
\caption{The real part of the Bondi news at late times ($u=-68.64$) for
$\epsilon=10^{-6}$. The dimple visible in the angular profile is due to the
presence of higher harmonics, $\ell > 2$.}
\label{fig:dimple}
\end{figure}

Figures~\ref{fig:quad} and \ref{fig:dimple} illustrate the angular
behavior of the news at early time ($u=-96.63$) and at late time
($u=-68.64$), respectively. We plot the real part of the Bondi
news on a single stereographic patch for the case $\epsilon
=10^{-6}$. The graphs clearly show that at early times the news is
pure quadrupole, in agreement with the perturbative regime of the
close approximation. The nonlinearity of the problem subsequently
introduces higher harmonics in the news. These higher harmonics
are clearly visible at later times in the plots of the Bondi news
as a function on the sphere. The most notable feature is the
dimple seen in Fig.~\ref{fig:dimple} on the profile of the news at
the pole (the center of the stereographic patch).

\begin{figure}
\epsfig{file=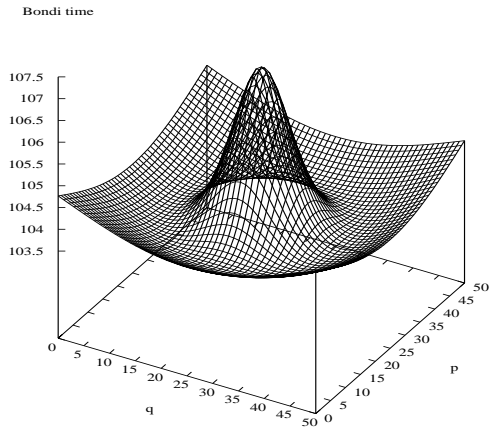,height=3.0in,width=3.0in,angle=0}
\caption{The Bondi time as a function on a stereographic patch at null
infinity at a late times ($u=-59.92$) for $\epsilon=10^{-6}$. The pole is
``running ahead'' of the equator, leading to the dimple observed in
Fig.~\protect\ref{fig:dimple}.}
\label{fig:bondi}
\end{figure}

This feature is partially due to the retardation effect introduced by the
angular-dependent redshift of the Bondi frames at future null infinity
${\cal I}^+$. We have chosen the initial cut of ${\cal I}^+$,
corresponding to a constant affine time $u$ on the horizon, to be a cut
of constant Bondi time. We then follow the inertial observers at ${\cal
I}^+$ to define a Bondi time slicing. Figure~\ref{fig:bondi}, which
displays the angular dependence of Bondi time $u_B$ for a fixed horizon
time $u$, reveals a pronounced increase in Bondi time at the pole relative
to the equator.

\section{Conclusions}
\label{sec:conclusion}

We have computed a family of spacetimes exterior to a head-on white
hole fission ranging from the close approximation to the nonlinear
regime. The results reveal a dramatic time and angular dependence
in the waveforms produced in the extreme nonlinear regime. At early
times, the results agree with close-approximation perturbative
calculations as expected. While the results presented here are for
the axisymmetric non-spinning case, the data and evolution codes
are not restricted to any symmetry. It will be interesting to see
how the results for a head-on collision are modified in the fission
of a spinning white hole.

Reexpressed in terms of the time-reversed scenario of a black hole
merger, the boundary conditions for a fission corresponds to no
{\it outgoing} radiation in the black hole case. Nevertheless, the
results pave the way for an application of characteristic codes to
calculate the fully nonlinear waveform emitted in a binary black
hole collision in the time period from merger to ringdown.
Waveforms from a black hole merger can be expected to differ from
those from a white hole fission, as has been observed in close
approximation studies~\cite{close2}. The fission process is
directly observable at ${\cal I}^+$ whereas the merger waveform
arises indirectly from the preceding collapse of the matter or
gravitational wave energy that forms the black holes. This
suggests that the fission is a more efficient source of
gravitational waves and that the high fractional mass losses
computed here cannot be attained in a black hole merger.

\begin{acknowledgements}

We thank Y.~Zlochower for helpful discussions and careful checking of the
numerical code. This research has been partially supported by NSF
grants PHY 9800731 and PHY 9988663 to the University of Pittsburgh,
and NSF grant PHY-0135390 to Carnegie Mellon University. L.~L. thanks
PIMS and CITA for support. R.~G. thanks the Albert Einstein Institute
for hospitality. Computer time for this project was provided by the
Pittsburgh Supercomputing Center.

\end{acknowledgements}

\end{document}